\documentclass[english]{article}
\usepackage[T1]{fontenc}
\usepackage[latin9]{inputenc}
\usepackage{geometry}
\usepackage{color}
\usepackage{booktabs}
\usepackage{amsmath}
\usepackage{graphicx}
\usepackage{subfigure,float}
\usepackage{esint}

\makeatletter

\providecommand{\tabularnewline}{\\}

\usepackage[sort&compress,numbers]{natbib}
\usepackage{times}

\date{}

\usepackage{babel}

\makeatother

\usepackage{babel}
\begin{document}
\title{Impact of photon addition and subtraction on nonclassical and phase
properties of a displaced Fock state }
\author{Priya Malpani$^{\dagger}$, Kishore Thapliyal$^{\ddagger}$, Nasir
Alam$^{\mathsection,*}$, Anirban Pathak$^{\mathsection}$, V. Narayanan$^{\dagger}$,\\
 Subhashish Banerjee$^{\dagger}$ \\
$^{\dagger}$Indian Institute of Technology Jodhpur, Jodhpur 342037,
India\\
$^{\ddagger}$RCPTM, Joint Laboratory of Optics of Palacky University
and Institute of Physics\\
of Academy of Science of the Czech Republic, Faculty of Science, Palacky
University, \\
 17. listopadu 12, 771 46 Olomouc, Czech Republic\\
$^{\mathsection}$Jaypee Institute of Information Technology, A-10,
Sector-62, Noida, UP-201309, India\\
$^{*}$Bankura University, Beside NH 60, Bankura, Block II, Purandarpur,
West Bengal 722155}
\maketitle
\begin{abstract}
Various nonclassical and quantum phase properties of photon added
then subtracted displaced Fock state have been examined systematically
and rigorously. Higher-order moments of the relevant bosonic operators
are computed to test the nonclassicality of the state of interest,
which reduces to various quantum states (having applications in quantum
optics, metrology and information processing) in different limits
ranging from the coherent (classical) state to the Fock (most nonclassical)
states. The nonclassical features are discussed using Klyshko's, Vogel's,
and Agarwal-Tara's criteria as well as the criteria of lower- and
higher-order antibunching, sub-Poissonian photon statistics and squeezing.
In addition, phase distribution function and quantum phase fluctuation
have been studied. These properties are examined for various combinations
of number of photon addition and/or subtraction and Fock parameter.
The examination has revealed that photon addition generally improves
nonclassicality, and this advantage enhances for the large (small)
values of displacement parameter using photon subtraction (Fock parameter).
The higher-order sub-Poissonian photon statistics is only observed
for the odd orders. In general, higher-order nonclassicality criteria
are found to detect nonclassicality even in the cases when corresponding
lower-order criteria failed to do so. Photon subtraction is observed
to induce squeezing, but only large number of photon addition can
be used to probe squeezing for large values of displacement parameter.
Further, photon subtraction is found to alter the phase properties
more than photon addition, while Fock parameter has an opposite effect
of the photon addition/subtraction. Finally, nonclassicality and non-Gaussianity
is also established using $Q$ function.
\end{abstract}

\section{Introduction \label{sec:Intro}}

As nonclassical states do not have any classical counterpart they
are essential for achieving quantum supremacy (\cite{douce2017continuous}
and references therein). The notion of nonclassicality used in the
present work is defined by the negative values of Glauber-Sudarshan
$P$ function which implies that a nonclassical state cannot be expressed
as a mixture of coherent states \cite{glauber1963coherent,sudarshan1963equivalence}.
Such states have been studied for long, but their importance has been
enhanced with the recent developments in quantum computation, communication
and sensing (see \cite{nielsen2002quantum,pathak2018classical,lanzagorta2011quantum}
and references therein). In fact, in the recent past, various exciting
applications of nonclassicality ranging from satellite based quantum
key distribution (QKD) \cite{liao2018satellite,sharma2019analysis}
to the detection of gravitational wave in LIGO \cite{aasi2013enhanced,grote2013first},
have been reported, and those have helped to establish that quantum
supremacy cannot be established without the use of nonclassical states
\cite{douce2017continuous}. Nonclassical states, in general, can
be created using various types of physical resources (e.g., PT symmetric
systems \cite{naikoo2019interplay,naikoo2019quantum}, quantum walk
\cite{banerjee2008symmetry,rao2011quantumness,srikanth2010quantumness},
atom-optical interactions \cite{naikoo2018probing,alam2019bose},
nonlinear optical couplers \cite{thapliyal2014higher,thapliyal2014nonclassical},
nonlinear optical processes \cite{thapliyal2019lower,thapliyal2019nonclassicality},
Bose-Einstein condensate \cite{giri2017nonclassicality}). We are
particularly interested in the realizations based on photonics, where
nonclassical states are generated by using linear and nonlinear optical
components, including mirrors, beam splitters, detectors, wave plates,
nonlinear crystals, pentaprism, beam displacer and retroreflectors
\cite{pathak2016optical}. The art of generating the required quantum
states is known as the quantum state engineering \cite{cooper2015characterization,dakna1998quantum,makhlin2001quantum,verstraete2009quantum},
which plays a very crucial role in generating various quantum states
which are required for the quantum information processing. 

There are some distinct theoretical tools for performing quantum state
engineering, like quantum scissoring \cite{miranowicz2001quantum},
hole-burning \cite{escher2004controlled,gerry2002hole,malpani2019filter}
or filtering out a particular Fock state from the photon number distribution
\cite{meher2018number}, applying non-Gaussianity inducing operations
\cite{agarwal2013quantum}. However, these distinct mechanisms are
realized primarily by appropriately using beam-splitter, mirror and
single photon detectors or single photon counting module. Without
going into finer details of optical realization of quantum state engineering
tools, we may note that these tools can be used to generate various
nonclassical states, e.g.,  displaced Fock state (DFS)\footnote{DFS is also referred to as generalized coherent state and displaced
number state (\cite{de1990properties,malpani2019lower} and references
therein).} \cite{de1990properties}, photon added DFS \cite{malpani2019lower},
photon subtracted DFS \cite{malpani2019lower}, photon added squeezed
coherent state \cite{thapliyal2017comparison}, photon subtracted
squeezed coherent state \cite{thapliyal2017comparison}, number state
filtered coherent state \cite{meher2018number}. Some of these states,
like photon added coherent state (PACS), have already been realized
experimentally \cite{zavatta2004quantum}.

Many of the above mentioned engineered quantum states have already
been studied in detail. Primarily, three types of investigations have
been performed on the engineered quantum states- (i) study of various
nonclassical features of these states (and their variation with the
state parameters) as reflected through different witnesses of nonclassicality.
Initially, such studies were restricted to the lower-order nonclassical
features. In the recent past, various higher-order nonclassical features
have been predicted theoretically \cite{alam2018higher,alam2018higher1,pathak2006control,pathak2010recent,verma2008higher}
and confirmed experimentally (\cite{hamar2014non,perina2017higher}
and references therein) in quantum states generated in nonlinear optical
processes. (ii) Phase properties of the nonclassical states have been
studied \cite{malpani2019quantum} by computing quantum phase fluctuations,
phase dispersion, phase distribution functions, etc., under various
formalisms, like Susskind and Glogower \cite{susskind1964quantum},
Pegg-Barnett \cite{pegg1989dt} and Barnett-Pegg \cite{barnett1986sm}
formalisms. (iii) Various applications of the engineered quantum states
have been designed. For example, antibunching is used for the characterization
of the single photon source \cite{pathak2006control}, squeezing (specifically,
squeezed vacuum) is used to achieve the required precision in measurement
in the LIGO experiment \cite{aasi2013enhanced,grote2013first}, and
entanglement is used in quantum teleportation \cite{bennett1993teleporting}
and quantum cryptography \cite{bennett1992quantumBBM}. More interestingly,
many of the engineered quantum states have been used in continuous
variable quantum cryptography (some of which will be mentioned in
the next section) (\cite{borelli2016quantum,srikara2019continuous}
and references therein). 

Motivated by the above observations, in what follows, we would like
to perform an investigation on a particularly interesting engineered
quantum state which would have the flavor of all the three facets
of studies mentioned above. Specifically, in what follows, we aim
to study the nonclassical (both lower- and higher-order) and phase
properties of a photon added then subtracted DFS (PASDFS) which can
be obtained by applying non-Gaussianity inducing operators on DFS.
The reason behind selecting this particular state lies in the fact
that this is a general state in the sense that in the limiting cases,
this state reduces to different quantum states having known applications
in continuous variable quantum cryptography (this point will be further
elaborated in the next section). 

As it appears from the above discussion, this investigation has two
facets. Firstly, we wish to study nonclassical features of PASDFS,
namely Klyshko's \cite{klyshko1996observable}, Agarwal-Tara's \cite{agarwal1992nonclassical},
Vogel's \cite{shchukin2005nonclassical} criteria, lower- and higher-order
antibunching \cite{pathak2006control}, squeezing \cite{hillery1987amplitude,hong1985generation,hong1985higher},
and sub-Poissonian photon statistics (HOSPS) \cite{zou1990photon}.
We subsequently study the phase properties of PASDFS by computing
phase distribution function \cite{agarwal1996complementarity,beck1993experimental},
phase fluctuation parameters \cite{barnett1986sm,carruthers1968phase},
and phase dispersion \cite{perinova1998phase}. A detailed analysis
of the obtained results will also be performed to reveal the usefulness
of the obtained results. 

The rest of the paper is organized as follows. In the next section,
we describe the quantum state of interest (i.e., PASDFS) in Fock basis
and calculate the analytic expressions for the higher-order moments
of the relevant field operators for this state. In Section \ref{sec:Nonclassicality-witnesses},
we investigate the possibilities of witnessing various nonclassical
features in PASDFS and its limiting cases by using a set of moments-based
criteria for nonclassicality. Variations of nonclassical features
(witnessed through different criteria) with various physical parameters
are also discussed here. In Section \ref{sec:Phase-properties-of},
phase properties of PASDFS are studied. $Q$ function for PASDFS is
obtained in Section \ref{sec:Qfn}. Finally, we conclude in Section
\ref{sec:Conclusions}.

\section{Quantum states of our interest \label{sec:Quantum-states-of}}

As mentioned in the previous section, this paper is focused on PASDFS.
Before, expressing PASDFS in Fock basis, we may note that a DFS can
be prepared by applying displacement operator on Fock state $|n\rangle$,
and the same can be expanded in Fock basis as 

\begin{equation}
|\phi(n,\alpha)\rangle=\hat{D}(\alpha)|n\rangle=\sum\limits _{m=0}^{\infty}C_{m}(\alpha,n)|m\rangle,\label{eq:GCS}
\end{equation}
where $\ensuremath{C_{m}(\alpha,n)=\left\langle m\left|\hat{D}(\alpha)\right|n\right\rangle }$.
A PASDFS can be obtained by sequentially applying appropriate number
of annihilation (photon subtraction) and creation (photon addition)
operators on a DFS. Analytical expression for PASDFS (specifically,
a $k$ photon added and then $q$ photon subtracted DFS) in Fock basis
can be shown to be 

\begin{equation}
|\psi(k,q,n,\alpha)\rangle=N\hat{a}^{q}\hat{a}^{\dagger k}|\psi(n,\alpha)\rangle=N\sum\limits _{m=0}^{\infty}C_{m}\left(\alpha,n,k,q\right)\left|m+k-q\right\rangle ,\label{eq:PADFS}
\end{equation}
where $N=\left[\sum\limits _{m=0}^{\infty}\left|C_{m}\left(\alpha,n,k,q\right)\right|^{2}\right]^{-\frac{1}{2}}$
is the normalization factor. A bit of computation yields the expression
for higher-order moment of annihilation and creation operator as 

\begin{equation}
\begin{array}{lcl}
\langle\hat{a}^{\dagger t}\hat{a}^{j}\rangle\equiv\langle\psi(k,q,n,\alpha)|\hat{a}^{\dagger t}\hat{a}^{j}|\psi(k,q,n,\alpha)\rangle & = & N^{2}\sum\limits _{m=0}^{\infty}C_{m}^{*}\left(\alpha,n,k,q\right)C_{m-j+t}\left(\alpha,n,k,q\right)\frac{\sqrt{\left(m+k-q\right)!\left(m+k-q-j+t\right)!}}{\left(m+k-q-j\right)!}.\end{array}\label{eq:PA-expepectation}
\end{equation}
For different values of $t$ and $j$, moments of any order can be
obtained, and the same may be used to investigate the nonclassical
properties of PASDFS and its limiting cases by using various moments-based
criteria of nonclassicality. The same will be performed in the following
section, but before proceeding, it would be apt to briefly state our
motivation behind the selection of this particular state for the present
study (or why do we find this state as interesting?).

Due to the difficulty in realizing single photon on demand sources,
the unconditional security promised by various QKD schemes, like BB84
\cite{bennett1984quantum} and B92 \cite{bennett1992quantum}, does
not remain unconditional in the practical situations This is where
continuous variable QKD (CVQKD) becomes relevant as they do not require
single photon sources. Special cases of PASDFS has already been found
useful in the realization of CVQKD. For example, protocols for CVQKD
have been proposed using PACS ($k=1,\,q=0,\,n=0$) \cite{pinheiro2013quantum,wang2014quantum},
photon added then subtracted coherent states ($k=1,\,q=1,\,n=0$)
\cite{borelli2016quantum,srikara2019continuous}, and coherent state
($k=0,\,q=0,\,n=0$) \cite{grosshans2002continuous,hirano2017implementation,huang2016long,ma2018continuous}.
Further, boson sampling with displaced single photon Fock states and
single PACS \cite{seshadreesan2015boson} has been reported, and an
$m$ PACS ($k=m,\,q=0,\,n=0$) has been used for quantum teleportation
\cite{pinheiro2013quantum}. Apart from these schemes of CVQKD, which
can be realized by using PASDFS or its limiting cases, the fact that
the photon addition and/or subtraction operation from a classical
or nonclassical state can be performed experimentally using the existing
technology \cite{parigi2007probing,zavatta2004quantum} has enhanced
the importance of PASDFS.

\section{Nonclassicality witnesses and the nonclassical features of PASDFS
witnessed through those criteria \label{sec:Nonclassicality-witnesses}}

The negative value of the Glauber-Sudarshan $P$ function characterizes
nonclassicality of an arbitrary state \cite{glauber1963coherent,sudarshan1963equivalence}.
As $P$ function is not directly measurable in experiments, many witnesses
of nonclassicality have been proposed, such as, negative values of
Wigner function \cite{wigner1932quantum,kenfack2004negativity}, zeroes
of $Q$ function \cite{husimi1940some,lutkenhaus1995nonclassical},
several moments-based criteria \cite{miranowicz2010testing,naikoo2018probing}.
An infinite set of such moments-based criteria of nonclassicality
is equivalent to $P$ function in terms of necessary and sufficient
conditions to detect nonclassicality \cite{richter2002nonclassicality}.
Here, we discuss some of these moments-based criteria of nonclassicality
and $Q$ function (in Section \ref{sec:Qfn}) to study nonclassical
properties of the state of our interest.

\subsection{Lower- and higher-order antibunching}

Nonclassicality witnessed through this criterion ensures that the
state under consideration is a suitable choice to be used as single
photon source \cite{agarwal2013quantum,pathak2010recent}. The criteria
of lower- and higher-order antibunching is given in terms of moments
of number operator \cite{pathak2006control}

\begin{equation}
d(l-1)=\langle\hat{a}^{\dagger l}\hat{a}^{l}\rangle-\langle\hat{a}^{\dagger}\hat{a}\rangle^{l}<0.\label{eq:HOA}
\end{equation}
Special case $d(1)$ corresponds to lower-order antibunching, while
$l>2$ represents $\left(l-1\right)$th higher-order antibunching.
The higher-order nonclassicality criteria are shown to be good witnesses
of weak nonclassicality \cite{avenhaus2010accessing,allevi2012measuring,thapliyal2014higher,thapliyal2017comparison}.

\begin{figure}
\begin{centering}
\subfigure[]{\includegraphics[scale=0.5]{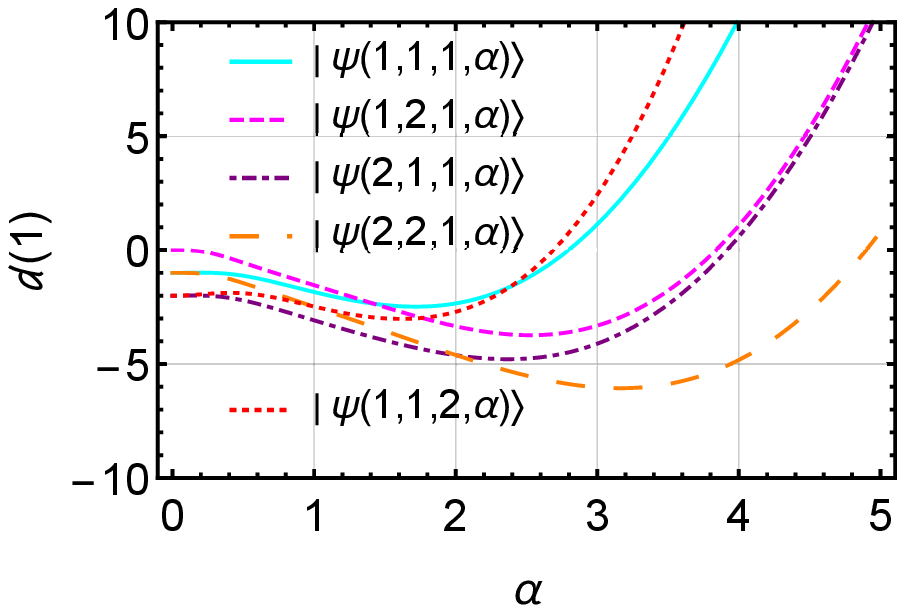}}  \subfigure[]{\includegraphics[scale=0.5]{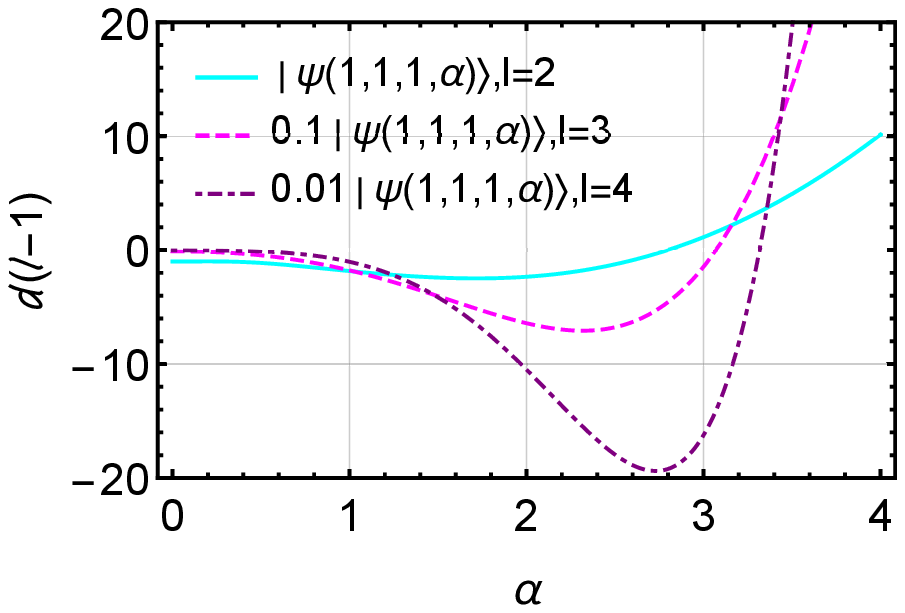}}\\ \subfigure[]{\includegraphics[scale=0.5]{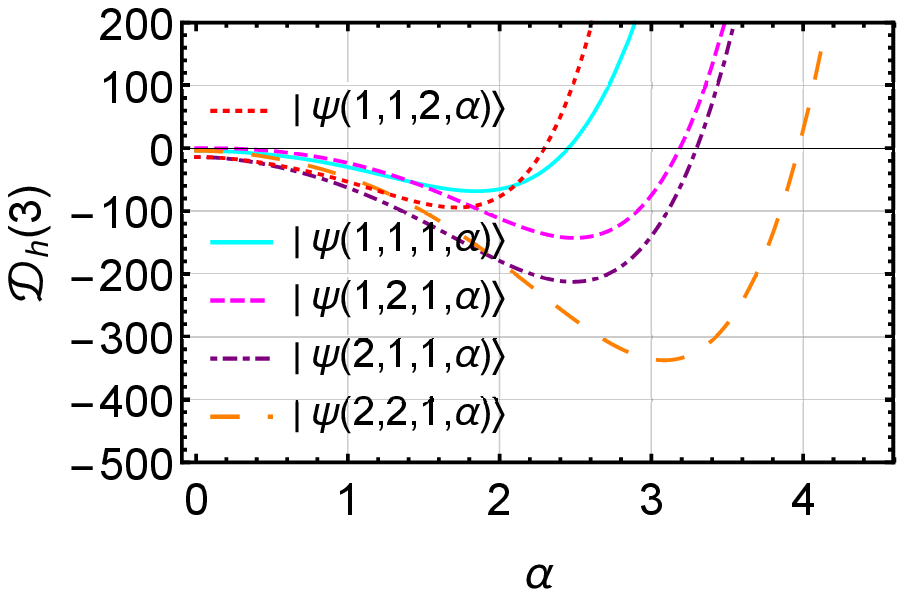}} \subfigure[]{\includegraphics[scale=0.5]{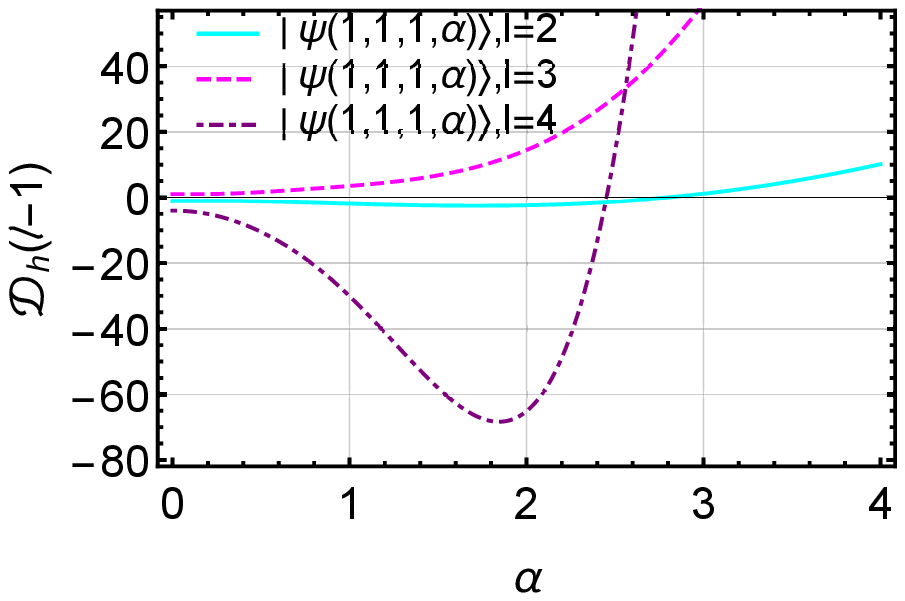}}
\par\end{centering}
\caption{\label{fig:HOSPS}(Color online) For PASDFS the lower-and higher-order
antibunching is given as a function of displacement parameter $\alpha$.
(a)\textcolor{green}{{} }\textcolor{black}{Lower-order antibunching
for }different values of parameters of the state. (b)\textcolor{green}{{}
}\textcolor{black}{Higher-order antibunching for particular state.}
HOSPS for PASDFS for \textcolor{black}{different }values of (c)\textcolor{green}{{}
}state parameters and (d)\textcolor{green}{{} }\textcolor{black}{order
of nonclassicality.}}
\end{figure}

The relevance of photon addition, photon subtraction, Fock, and displacement
parameters in the nonclassical properties of the class of PASDFSs
is studied here rigorously. Specifically, using Eq. (\ref{eq:PA-expepectation})
with the criterion of antibunching (\ref{eq:HOA}) we can study the
possibilities of observing lower- and higher-order antibunching in
the quantum states of PASDFS class, where the class of PASDFSs refers
to all the states that can be reduced from state (\ref{eq:PADFS})
in the limiting cases. The outcome of such a study is illustrated
in Fig. \ref{fig:HOSPS}. It is observed that the depth of lower-
and higher-order nonclassicality witnesses can be increased by increasing
the value of the displacement parameter, but large values of $\alpha$
deteriorate the observed nonclassicality (cf. Fig. \ref{fig:HOSPS}
(a)-(b)). The nonclassicality for higher-values of displacement parameter
$\alpha$ can be induced by subtracting photons at the cost of reduction
in the depth of nonclassicality witnessed for smaller $\alpha$, as
shown in Fig. \ref{fig:HOSPS} (a). However, photon addition is always
more advantageous than subtraction. Therefore, both addition and subtraction
of photons illustrate these collective effects by showing nonclassicality
for even higher values of $\alpha$ at the cost of that observed for
the small values of displacement parameter. Fock parameter has completely
opposite effect of photon subtraction as it shows the advantage (disadvantage)
for small (large) values of displacement parameter. Figure \ref{fig:HOSPS}
(b) shows benefit of studying higher-order nonclassicality as depth
of corresponding witness of nonclassicality can be observed to increase
with the order. The higher-order nonclassicality criterion is also
able to detect nonclassicality for certain values of displacement
parameter for which the corresponding lower-order criterion failed
to do so.

\subsection{Higher-order sub-Poissonian photon statistics}

Lower-order sub-Poissonian photon statistics is closely associated
with lower-order antibunching (\ref{eq:HOA}) criteria (\cite{thapliyal2017comparison}
and references therein). However, the presence of corresponding higher-order
nonclassical feature, i.e., HOSPS, is independent of higher-order
antibunching. HOSPS characterizes to the reduction of higher-order
variance of the number operator calculated for certain quantum state
with respect to the coherent state. The criterion can be expressed
as \cite{thapliyal2017comparison,malpani2019lower}

\begin{equation}
\begin{array}{lcccc}
\mathcal{D}_{h}(l-1) & = & \sum\limits _{e=0}^{l}\sum\limits _{f=1}^{e}S_{2}(e,\,f)\,^{l}C_{e}\,\left(-1\right)^{e}d(f-1)\langle\hat{N}\rangle^{l-e} & < & 0,\end{array}\label{eq:hosps22}
\end{equation}
where $S_{2}(e,f)$ is Stirling number of second kind, and $^{l}C_{e}$
is the usual binomial coefficient. HOSPS in PASDFS can be studied
using Eq. (\ref{eq:PA-expepectation}) in Eq. (\ref{eq:hosps22}).
Variation of HOSPS nonclassicality witness for class of PASDFSs obtained
by different nonclassicality inducing operations show the same effect
as that of antibunching witness for all the odd orders of HOSPS, and
as depicted in Fig. \ref{fig:HOSPS} (c). However, this nonclassical
feature disappears for even orders of HOSPS (cf. Fig. \ref{fig:HOSPS}
(d)). In case of the odd orders of HOSPS, though the depth of nonclassicality
witness increases with the order, higher-order criterion is found
to fail to detect nonclassicality for certain values of $\alpha$
when corresponding HOSPS criterion for smaller values of orders shown
the nonclassicality.

\subsection{Lower- and higher-order squeezing}

The variance in quadrature below the corresponding value for displaced
vacuum state (coherent state) can be defined as the squeezing in that
quadrature \cite{loudon1987squeezed}. Higher-order counterparts of
this effect are studied in two ways Hong-Mandel-type \cite{hong1985generation,hong1985higher}
and Hillery's amplitude-powered \cite{hillery1987amplitude} squeezing.
Here, we focus on the Hong-Mandel-type squeezing, which focuses on
the reduction of higher-order variances of quadrature, that can be
defined as \cite{thapliyal2017comparison,malpani2019lower}

\begin{equation}
\begin{array}{lcl}
\mathcal{S}\left(l\right)=\frac{\langle(\Delta X)^{l}\rangle-\left(\frac{1}{2}\right)_{\frac{l}{2}}}{\left(\frac{1}{2}\right)_{\frac{l}{2}}} & < & 0\end{array},\label{eq:Hong-def2-1}
\end{equation}
where $(l)_{x}$ is Pochhammer symbol, and $l$th-order variance $\langle(\Delta X)^{l}\rangle$
can be written in terms of moments of annihilation and creation operators
as \cite{verma2008higher}
\begin{equation}
\langle\left(\text{\ensuremath{\Delta}}X\right)^{l}\rangle=\sum\limits _{r=0}^{l}\sum\limits _{i=0}^{\frac{r}{2}}\sum\limits _{k=0}^{r-2i}\left(-1\right)^{r}\frac{1}{2^{\frac{l}{2}}}\left(2i-1\right)!\,^{2i}C_{k}{}^{l}C_{r}{}^{r}C_{2i}\langle\hat{a}^{\dagger}+\hat{a}\rangle^{l-r}\langle\hat{a}^{\dagger k}\hat{a}^{r-2i-k}\rangle.\label{eq:cond2.1}
\end{equation}
Only even values of $l$ are allowed in the criterion of Hong-Mandel-type
higher-order squeezing. 

\begin{figure}
\begin{centering}
\subfigure[]{\includegraphics[scale=0.5]{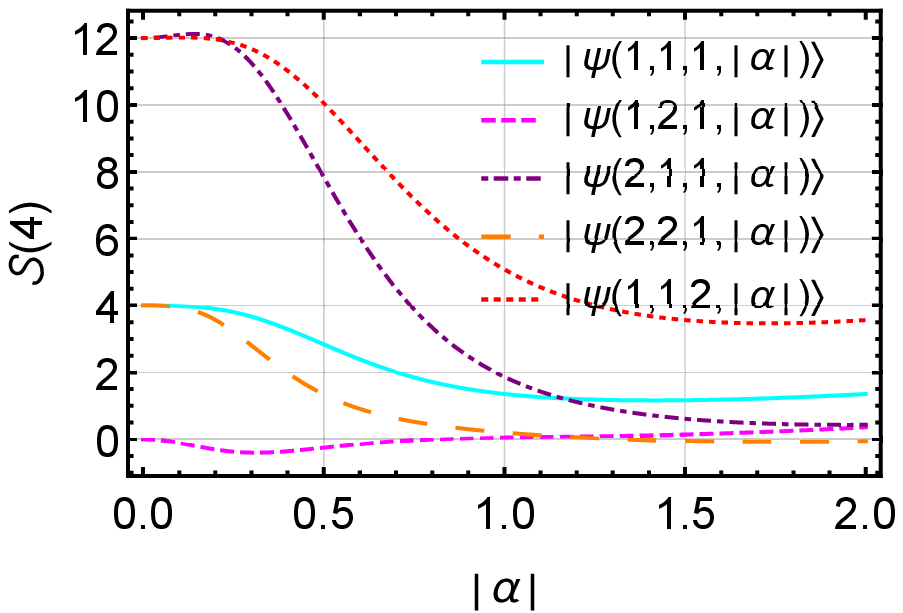}} \subfigure[]{\includegraphics[scale=0.5]{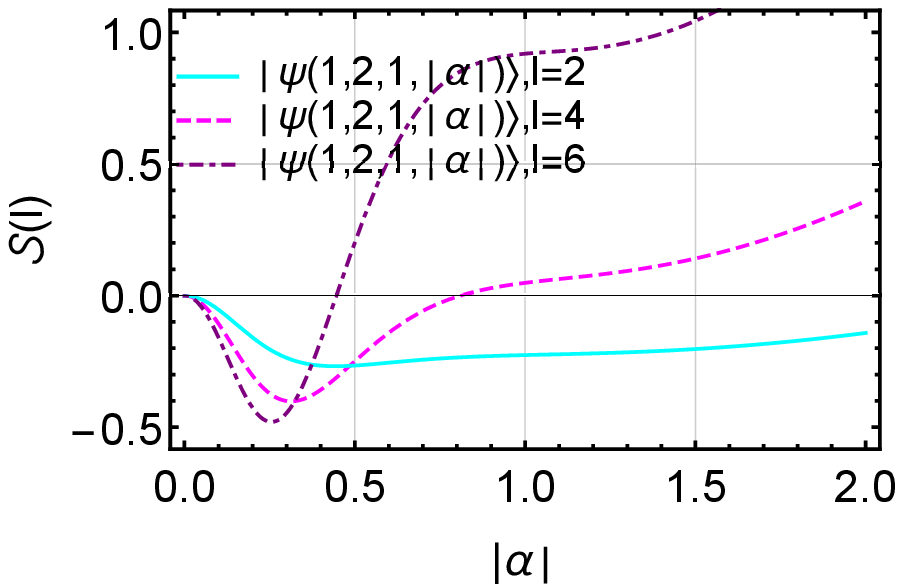}} \subfigure[]{\includegraphics[scale=0.5]{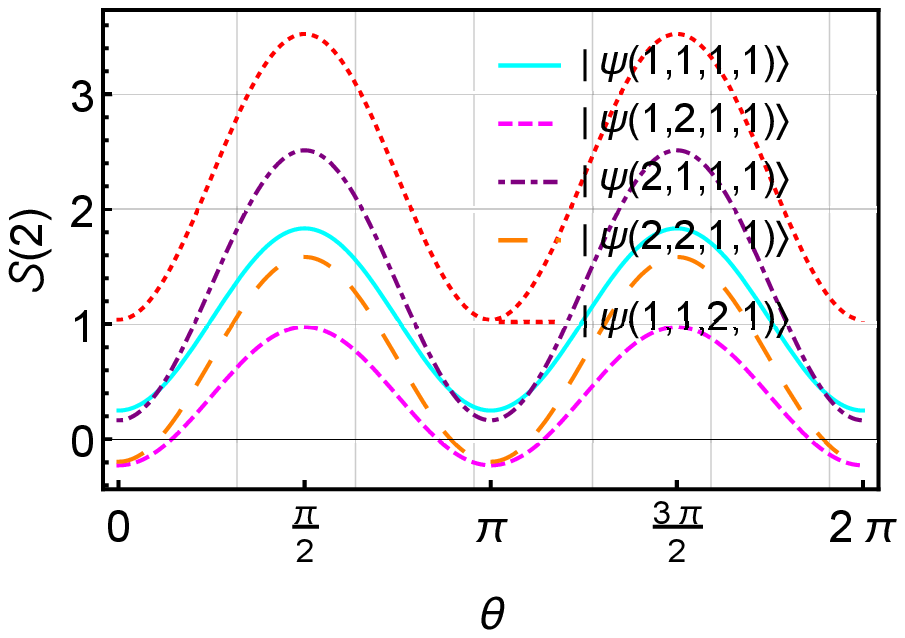}}
\par\end{centering}
\caption{\label{fig:HOS}(Color online) Dependence of the Hong-Mandel-type
higher-order squeezing witness on displacement parameter for (a) different
state parameters and (b) order of squeezing. (c) Lower-order squeezing
as a function of phase parameter of the state, i.e., phase $\theta$
of displacement parameter $\alpha=1.\exp[\iota\theta].$}
\end{figure}

Out of all the nonclassicality inducing operations used in PASDFS
only photon subtraction is squeezing inducing operation as shown in
Fig. \ref{fig:HOS}, which is consistent with some of our recent observations
\cite{malpani2019lower}. With photon addition higher-order squeezing
can be induced for large values of modulus of displacement parameter
at the cost of squeezing observed for small $\left|\alpha\right|$
as long as the number of photon subtracted is more than the value
of Fock parameter. As far as higher-order squeezing is concerned,
the observed nonclassicality disappears for large values of real displacement
parameter with increase in the depth of the nonclassicality witness.
Squeezing being a phase dependent nonclassical feature depends on
the phase $\theta$ of the displacement parameter $\alpha=\left|\alpha\right|\exp[\iota\theta]$
(shown in Fig. \ref{fig:HOS} (c) for lower-order squeezing). 

\begin{figure}
\begin{centering}
\subfigure[]{\includegraphics[scale=0.5]{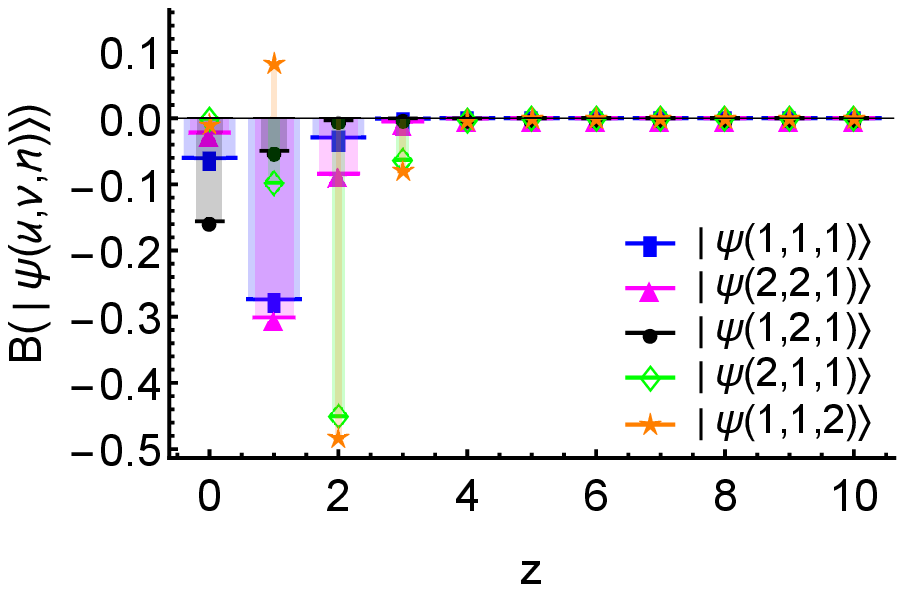}} \subfigure[]{\includegraphics[scale=0.5]{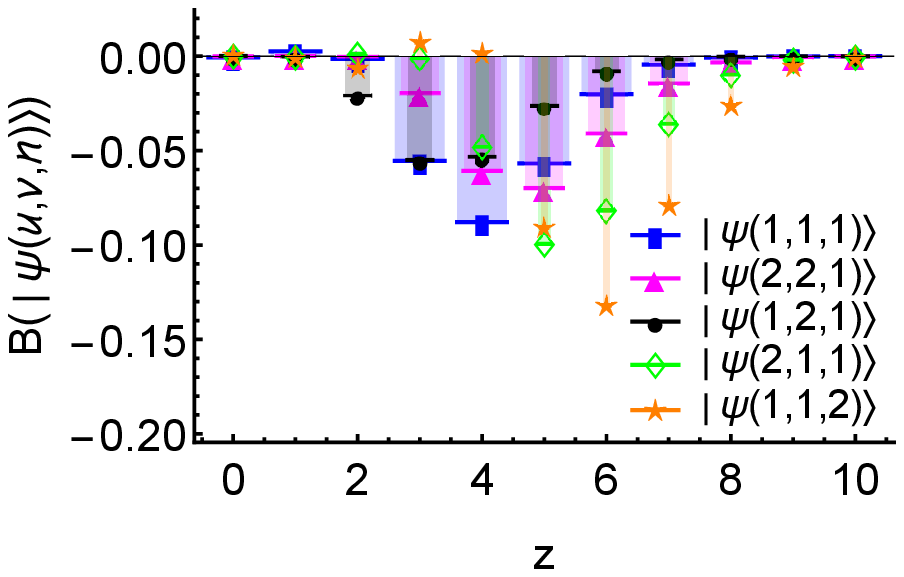}}
\par\end{centering}
\caption{\label{fig:Klyshko}\textcolor{green}{{} }(Color online) Illustration
of Klyshko's parameter $B\left(z\right)$ with respect to the photon
number $z$\textcolor{black}{{} for }different values of state parameters\textcolor{green}{{}
}with (a) $\alpha=0.5$ and (b) $\alpha=1$.}
\end{figure}

\subsection{Klyshko's Criterion}

Another interesting criterion of nonclassicality is based on the probability
of three successive photon numbers , i.e., $z,\,z+1,\,z+2$, and is
known as Klyshko's criterion \cite{klyshko1996observable}. It can
be defined in terms of probability $p_{z}$ of detecting $z$ number
of photons as

\begin{equation}
B(z)=(z+2)p_{z}p_{z+2}-(z+1)\left(p_{z+1}\right)^{2}<0.\label{eq:Klyshko}
\end{equation}
For PASDFS $p_{z}$ can be obtained from Eq. (\ref{eq:PA-expepectation}).
Nonclassicality reflected through Klyshko's criterion can be controlled
by all the state engineering operations used here as shown in Fig.
\ref{fig:Klyshko}. The depth of this nonclassicality witness increases
at higher values of photon numbers $z$ due to increase in photon
addition and/or Fock parameter. In contrast, depth of witness increases
at smaller photon numbers $z$ due to photon subtraction. The Klyshko's
nonclassicality witness is positive for some photon numbers only if
$k+n>q$. Additionally, with increase in displacement parameter the
depth of nonclassicality witness decreases, and the weight of the
distribution of witness shift to higher values of $z$.

\subsection{Agarwal-Tara's criterion}

\begin{figure}
\centering{}%
\subfigure[]{\includegraphics[scale=0.5]{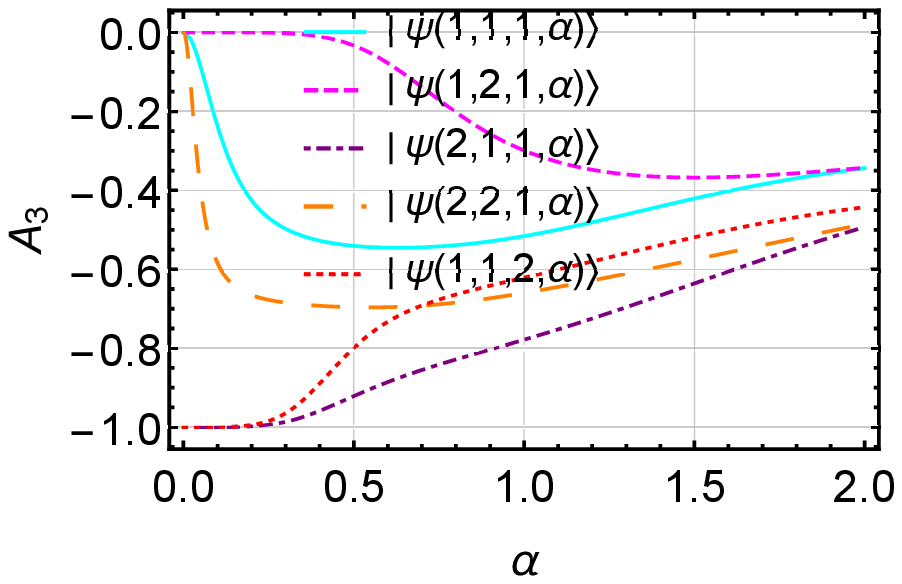}} \subfigure[]{\includegraphics[scale=0.5]{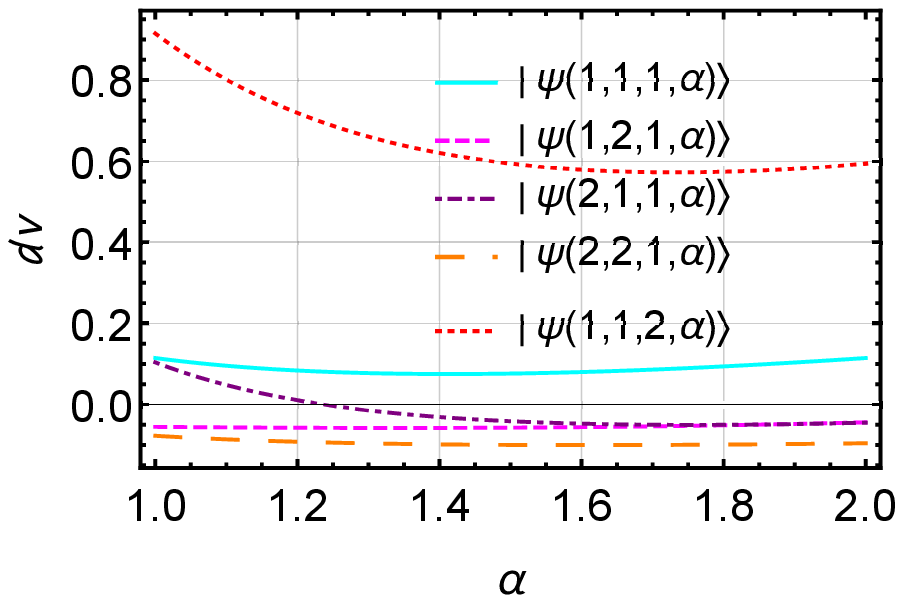}}\caption{\label{Vogel's criteria}(Color online) Nonclassicality reflected
through the negative values of (a) Agarwal-Tara's and (b) Vogel's
criteria as a function of $\alpha$ f\textcolor{black}{or different
state parameters.} }
\end{figure}

A moments-based criterion of nonclassicality was introduced in terms
of higher-order moments of number operator by Agarwal and Tara \cite{agarwal1992nonclassical}.
This criterion can be written as

\begin{equation}
A_{3}=\dfrac{\det m^{(3)}}{\det\mu^{(3)}-\det m^{(3)}}<0,\label{eq:Agarwal}
\end{equation}
where the matrices are given as 

\begin{equation}
s^{(3)}=\begin{bmatrix}1 & s_{1} & s_{2}\\
s_{1} & s_{2} & s_{3}\\
s_{2} & s_{3} & s_{4}
\end{bmatrix}\label{moment}
\end{equation}
with $s_{i}\in\left\{ m_{i},\mu_{i}\right\} $ and $m_{n}=\langle\hat{a}^{\dagger n}\hat{a}^{n}\rangle$
and $\mu_{n}=\langle\left(\hat{a}^{\dagger}\hat{a}\right)^{n}\rangle.$
This nonclassicality witness is able to detect nonclassicality in
all the quantum states in the class of PASDFSs (cf. Fig. \ref{Vogel's criteria}
(a)). Note that for $|\psi\left(1,2,1,\alpha\right)\rangle$ with
small $\alpha$, $A_{3}$ parameter is close to zero, which is due
to very high probability for zero photon states.

\subsection{Vogel's criterion}

The moments-based nonclassicality criterion of the previous subsection
was later extended to Vogel's nonclassicality criterion \cite{shchukin2005nonclassical}
in terms of matrix of moments as 

\begin{equation}
v=\left[\begin{array}{ccc}
1 & \langle\hat{a}\rangle & \langle\hat{a}^{\dagger}\rangle\\
\langle\hat{a}^{\dagger}\rangle & \langle\hat{a}^{\dagger}\hat{a}\rangle & \langle\hat{a}^{\dagger2}\rangle\\
\langle\hat{a}\rangle & \langle\hat{a}^{2}\rangle & \langle\hat{a}^{\dagger}\hat{a}\rangle
\end{array}\right].\label{eq:vogel}
\end{equation}
The negative value of the determinant $dv$ of matrix $v$ in Eq.
(\ref{eq:vogel}) is signature of nonclassicality. Fock parameter
has adverse effect on the nonclassicality in PASDFS detected by this
criterion. This averse effect can be compensated by photon subtraction
and can be further controlled by photon addition (as shown in Fig.
\ref{Vogel's criteria} (b)). Notice that the nonclassical behavior
illustrated by Agarwal-Tara's (Vogel's) criterion is related to higher-order
antibunching (squeezing) criterion. However, nonclassicality witness
of Vogel's criterion is a phase independent property unlike squeezing.

\section{Phase properties of PASDFS\label{sec:Phase-properties-of}}

The nonclassicality inducing operations are also expected to impact
the phase properties of a quantum state \cite{banerjee2007phase,banerjee2007phase1}.
Recently, we have reported an extensive study on the role that such
quantum state engineering tools can play in application oriented studies
on quantum phase \cite{malpani2019quantum}. Specifically, relevance
in quantum phase estimation, phase fluctuation, and phase distribution
were discussed which can play an important role in quantum metrology
\cite{giovannetti2011advances}. Here, we briefly discuss some of
the phase properties of the class of PASDFSs.

\subsection{Phase distribution function}

Phase distribution function for an arbitrary density operator $\varrho$
is defined as \cite{agarwal1992classical}

\begin{equation}
P(\theta)=\frac{1}{2\pi}\langle\theta|\varrho|\theta\rangle,\label{eq:Phase-Distridution}
\end{equation}
where phase state $|\theta\rangle$ is 

\begin{equation}
|\theta\rangle=\sum_{n=0}^{\infty}\exp^{\iota n\theta}|n\rangle.\label{phase-state}
\end{equation}
Fock states, used here for preparation of PASDFS, have uniformly distributed
phase $P_{\theta}=\frac{1}{2\pi}$. Further, photon addition and subtraction
are also used in PASDFS after applying displacement operator on Fock
state. The analytical expression for phase distribution function for
PASDFS can be computed as

\begin{equation}
\begin{array}{lcl}
P(\theta) & = & \frac{N^{2}}{2\pi}\left|\sum\limits _{m=0}^{\infty}C_{m}\left(\alpha,n,k,q\right)\exp\left[\iota\left(m+k-q\right)\theta\right]\right|^{2}.\end{array}\label{eq:PA-phase}
\end{equation}

\begin{figure}
\begin{centering}
\subfigure[]{\includegraphics[scale=0.4]{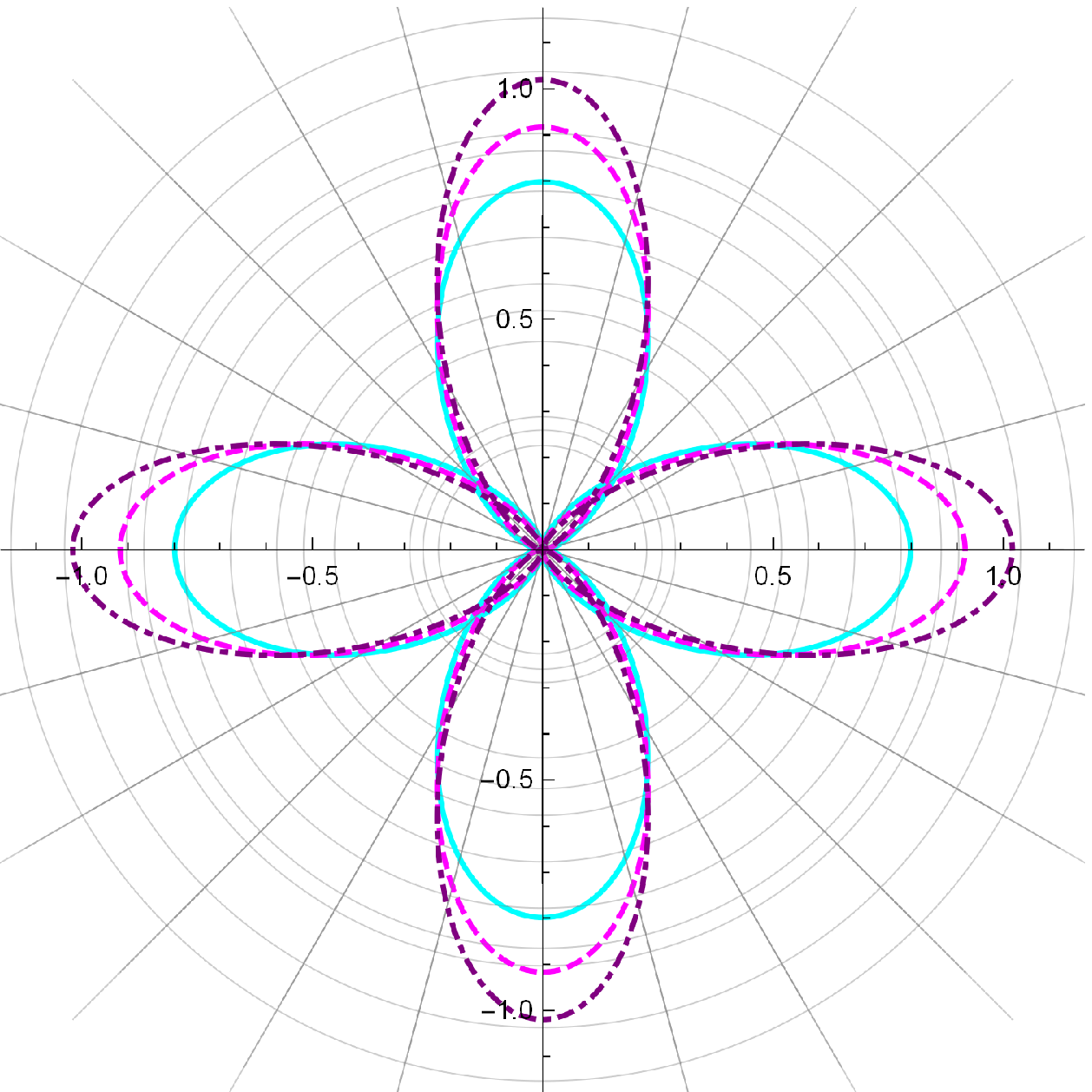}} \subfigure[]{\includegraphics[scale=0.4]{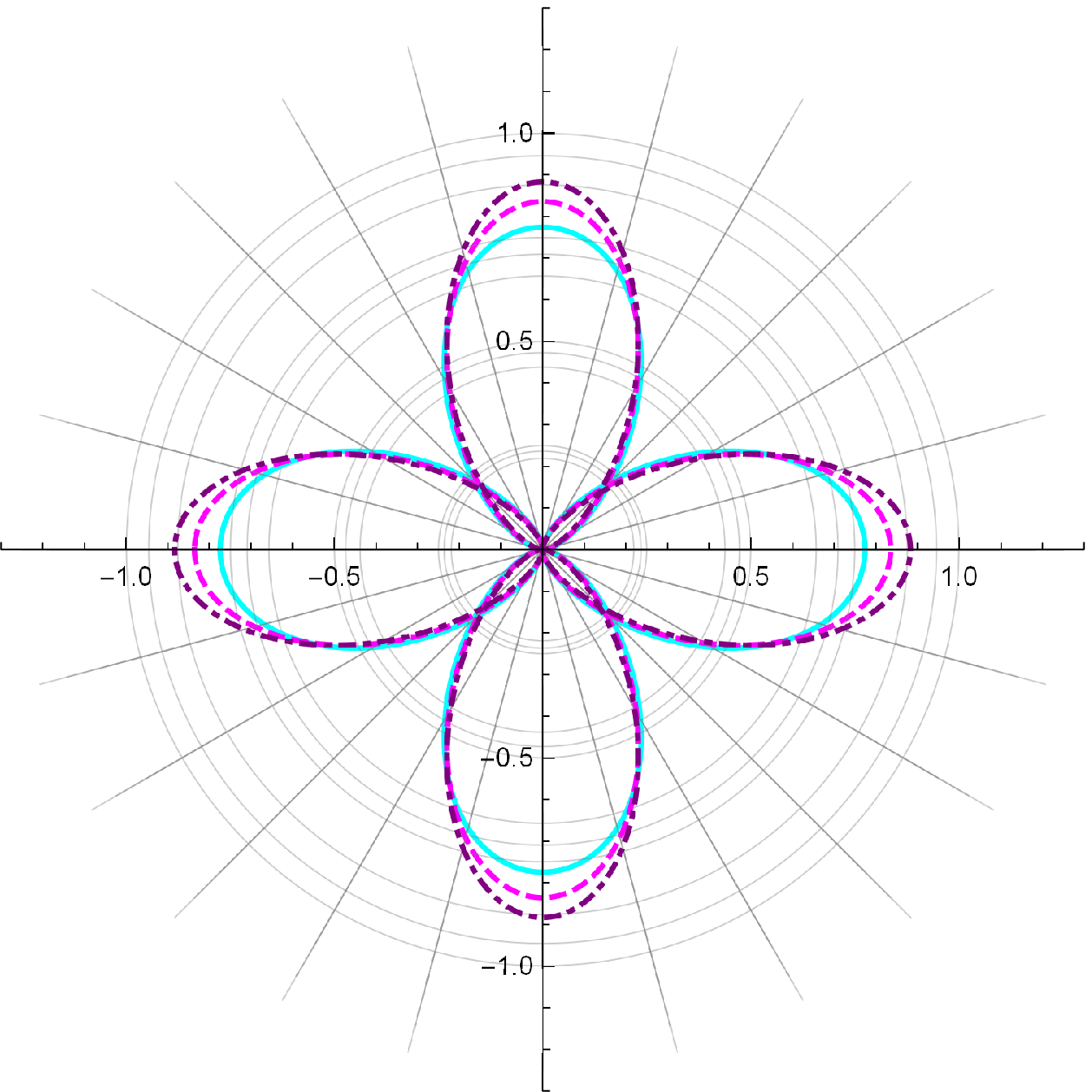}} \subfigure[]{\includegraphics[scale=0.4]{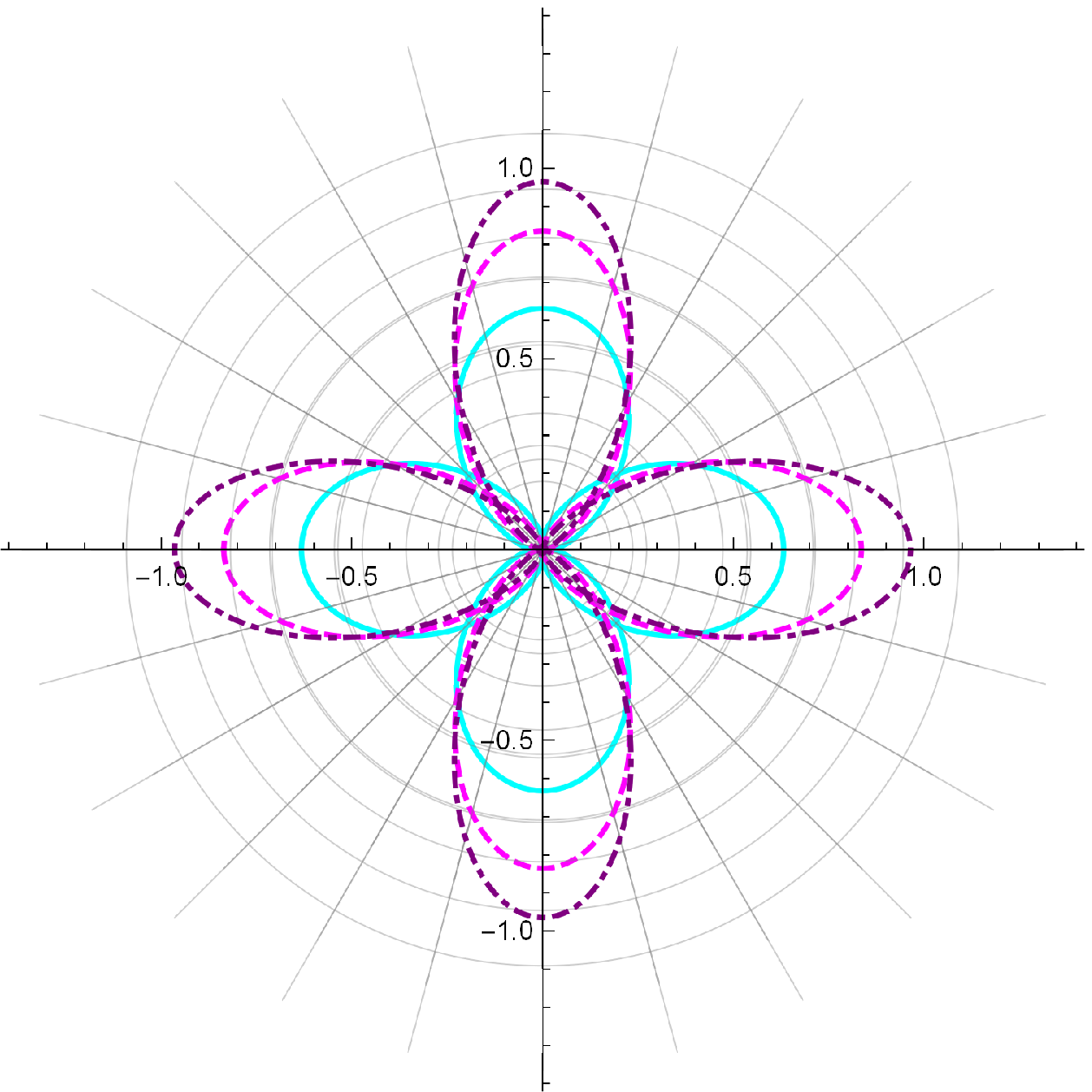}}
\par\end{centering}
\caption{\label{fig:Phase distribution function} (Color online) Polar plot
of phase distribution function for PASDFS $|\psi\left(k,q,n,\alpha\right)\rangle$
with respect to variation in displacement parameter for (a) $n=1,\,k=2$
and $q=1,$ 2, and 3 represented by the smooth (cyan), dashed (magenta),
and dot-dashed (purple) lines, respectively; (b) $n=2,\,q=2$ and
$k=1,$ 2, and 3 illustrated by the smooth (cyan), dashed (magenta),
and dot-dashed (purple) lines, respectively; and (c) $n=1$ with $k=q=1,$
2, and 3 shown by the smooth (cyan), dashed (magenta), and dot-dashed
(purple) lines, respectively.}
\end{figure}

Photon subtraction can be observed to be a more effective tool to
alter phase properties of PASDFS than photon addition, as shown in
Fig. \ref{fig:Phase distribution function}. Interestingly, photon
addition shows similar behavior, though less prominent, as photon
subtraction, Fock parameter has opposite effect. 

\begin{figure}
\begin{centering}
\includegraphics[scale=0.5]{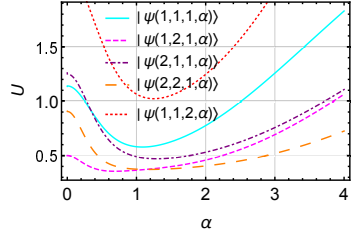}
\par\end{centering}
\caption{\label{fig:Phase fluctuation}\textcolor{green}{{} }(Color online) Variation
of phase fluctuation parameter with displacement parameter\textcolor{green}{{}
}for various state parameters in PASDFS.}
\end{figure}

\subsection{Phase Fluctuation}

The idea of phase operator was first given by Dirac \cite{dirac1927quantum}
with the assumption that $\hat{a}$ can be written as multiplication
of a unitary operator and a Hermitian function of $\hat{N}$ but it
led to an uncertainty relation that lacked physical meaning. Later,
Louisell came up with an idea of periodic phase \cite{louisell1963amplitude},
following which Susskind and Glogower developed the Sine and Cosine
operators \cite{susskind1964quantum}, which was further modified
by Barnett and Pegg \cite{barnett1986sm} as

\begin{equation}
\hat{S}=\frac{1}{2\iota}\left[\frac{1}{\left(\bar{N}+0.5\right)^{\frac{1}{2}}}\hat{a}-\hat{a}^{\dagger}\frac{1}{\left(\bar{N}+0.5\right)^{\frac{1}{2}}}\right]\label{eq:Phase fluctuation 1}
\end{equation}
and

\begin{equation}
\hat{C}=\frac{1}{2}\left[\frac{1}{\left(\bar{N}+0.5\right)^{\frac{1}{2}}}\hat{a}+\hat{a}^{\dagger}\frac{1}{\left(\bar{N}+0.5\right)^{\frac{1}{2}}}\right].\label{eq:Phase fluctuation 2}
\end{equation}
Here, $\bar{N}$ is the mean number of photons. Carruthers and Nieto
\cite{carruthers1968phase} provided phase fluctuation parameters
in terms of these operators

\begin{equation}
U=\left(\Delta N\right)^{2}\left[\left(\Delta S\right)^{2}+\left(\Delta C\right)^{2}\right]/\left[\langle\hat{S}\rangle^{2}+\langle\hat{C}\rangle^{2}\right],\label{eq:Phase fluctuation 3}
\end{equation}
and $S_{s}=\left(\Delta N\right)^{2}\left(\Delta S\right)^{2},\ Q=S_{s}/\langle\hat{C}\rangle^{2}.$
Here, we focus only on the first phase fluctuation parameter $U$,
which is related to antibunching if $U$ is below its value for coherent
state (i.e., 0.5), remaining consistent with Barnett-Pegg formalism
\cite{gupta2007reduction,pathak2000phase}. One can observe that the
phase fluctuation parameter is able to detect nonclassicality (specifically
antibunching) only in three cases where the role of the photon subtraction
is relevant (cf. Fig. \ref{fig:Phase fluctuation}). The observation
can be seen analogous to that observed for Vogel's nonclassicality
criterion.

\begin{figure}
\begin{centering}
\subfigure[]{\includegraphics[scale=0.3]{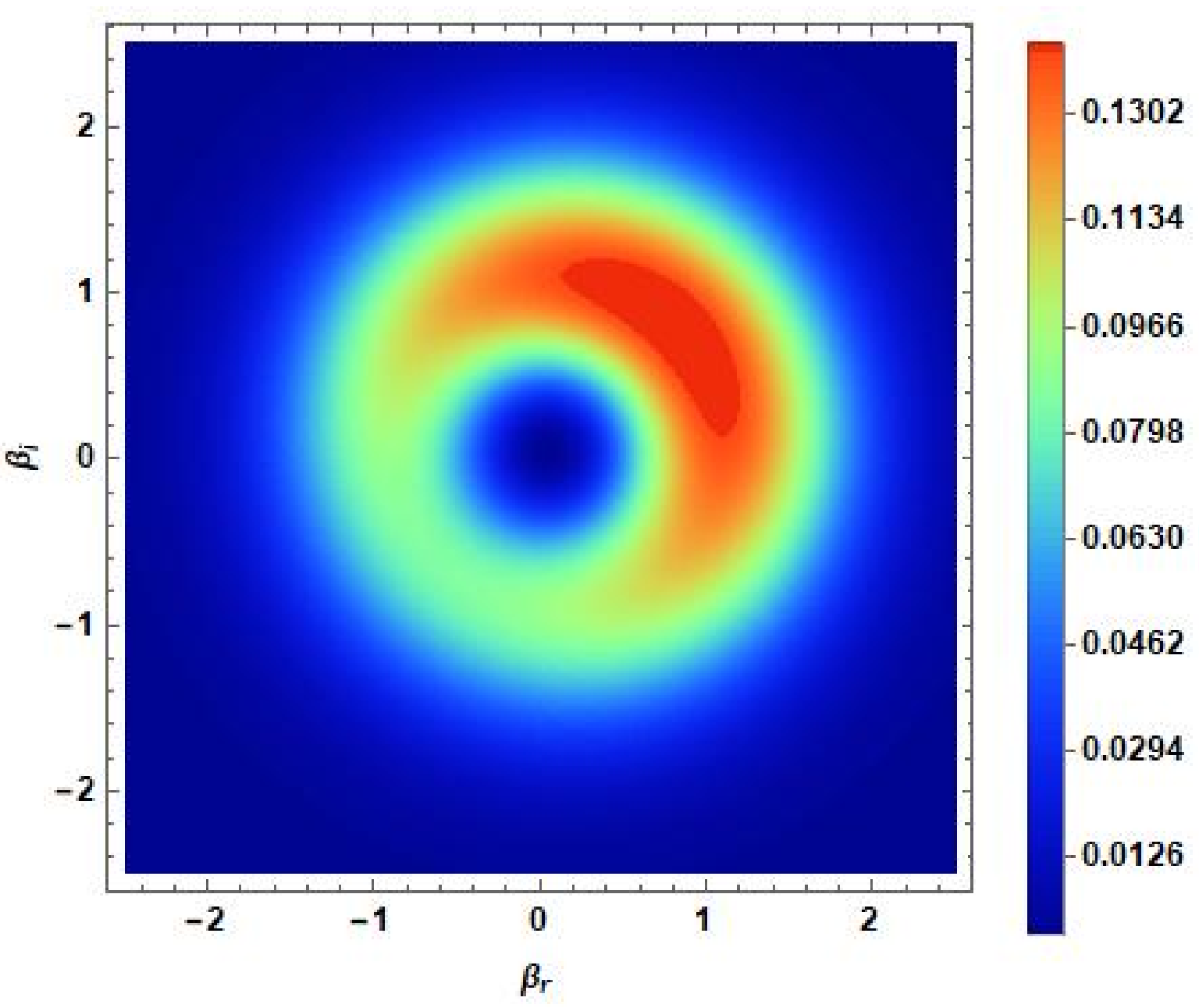}} \subfigure[]{\includegraphics[scale=0.3]{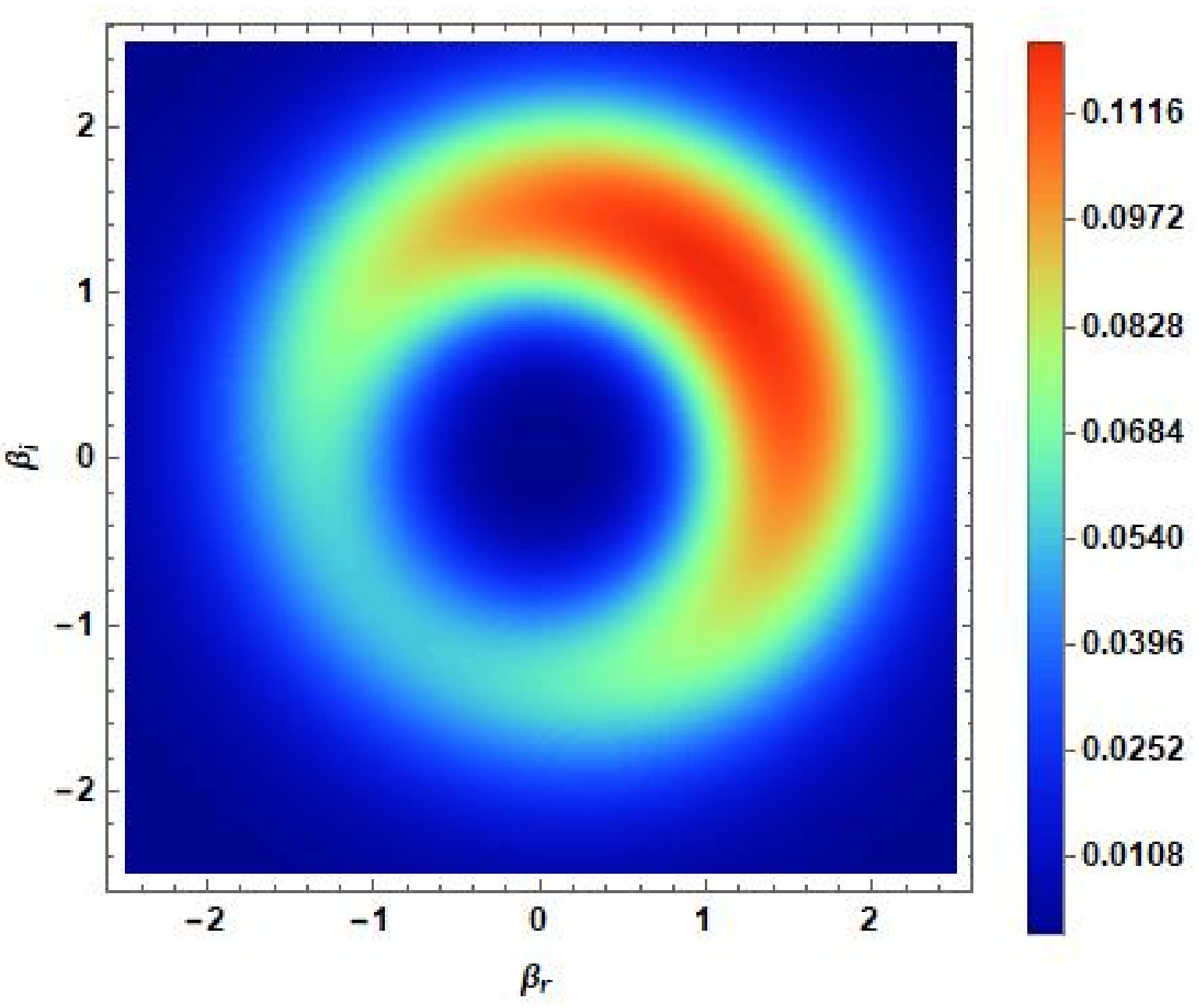}}\\
\subfigure[]{\includegraphics[scale=0.3]{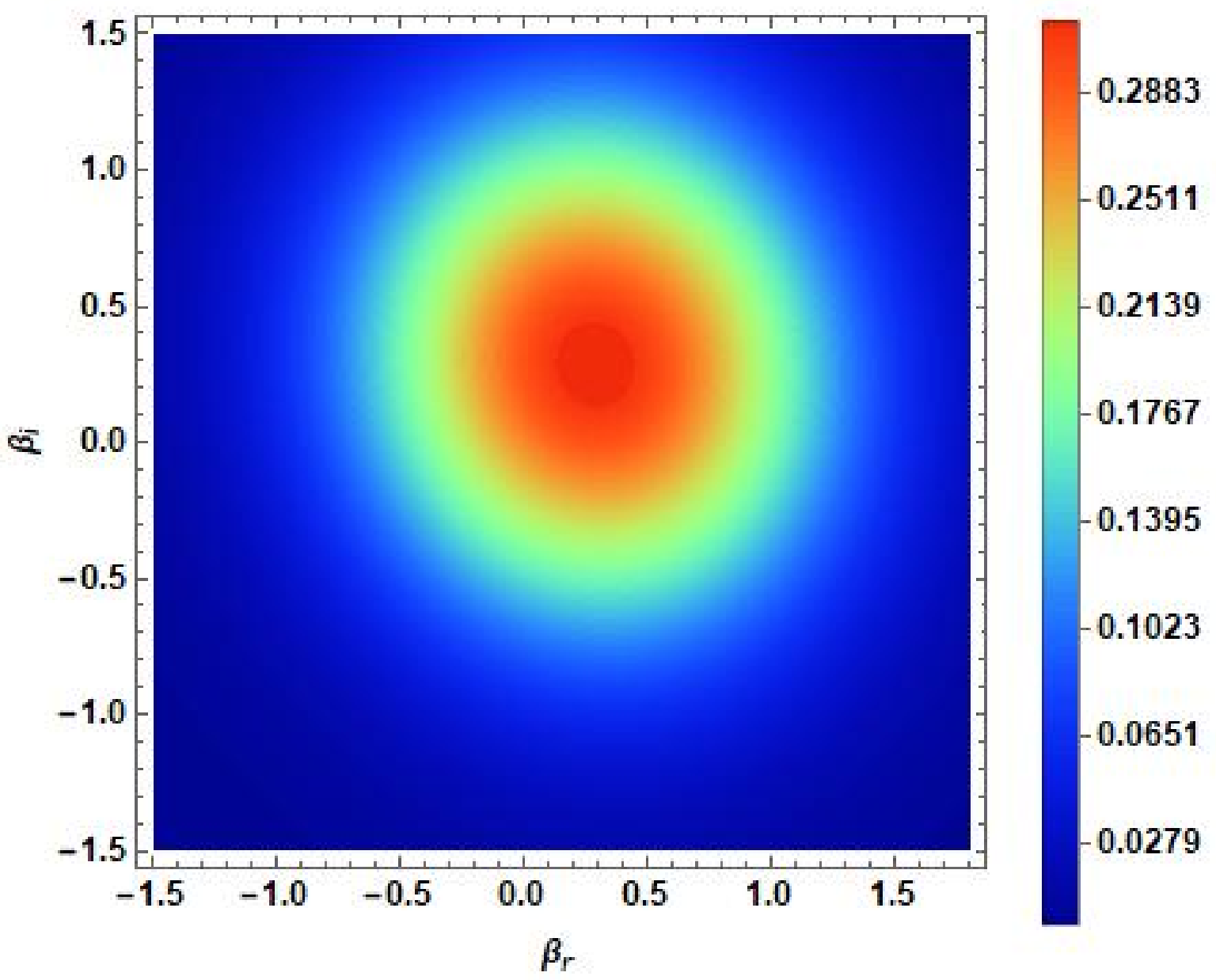}} \subfigure[]{\includegraphics[scale=0.3]{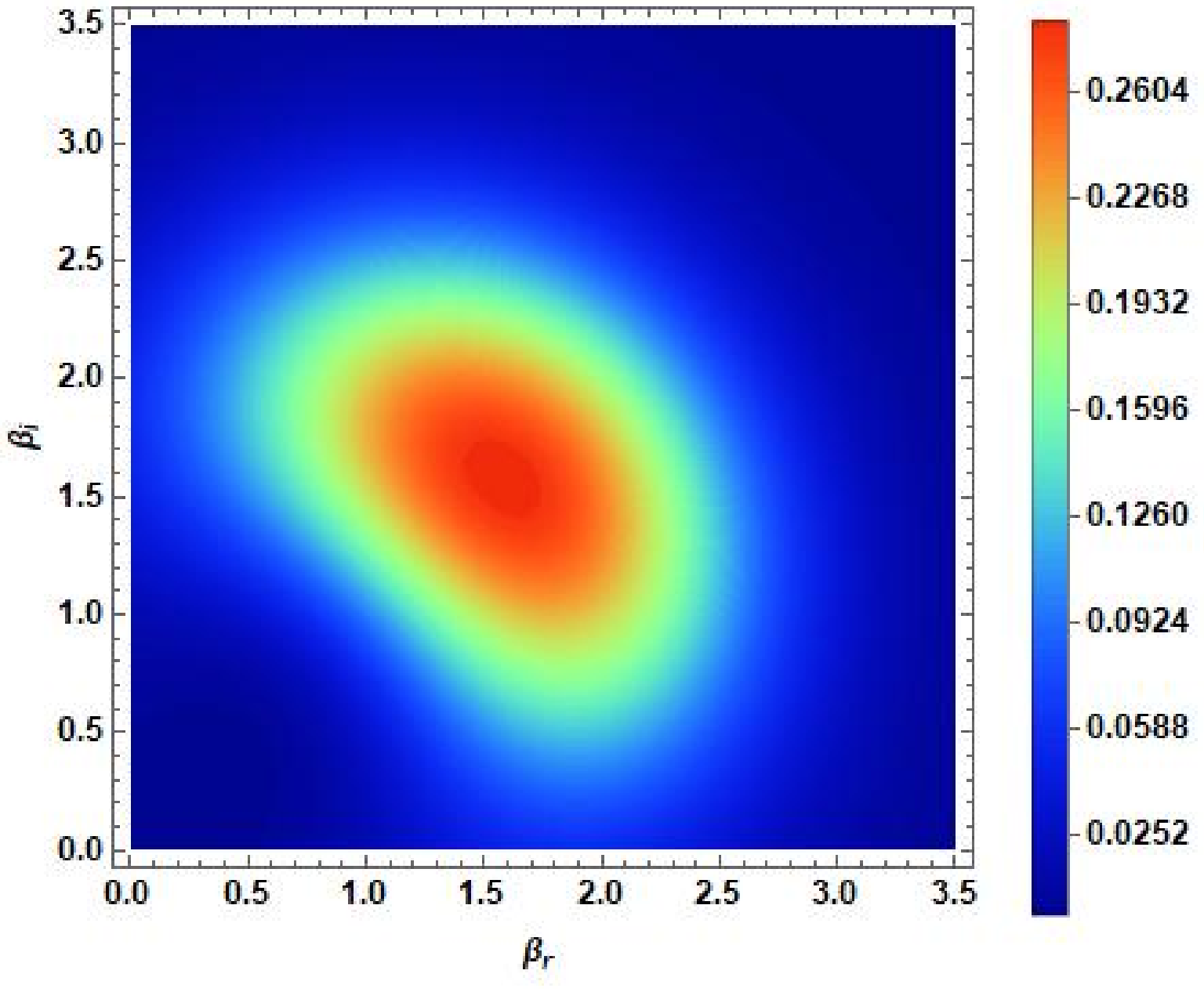}}\\
\subfigure[]{\includegraphics[scale=0.3]{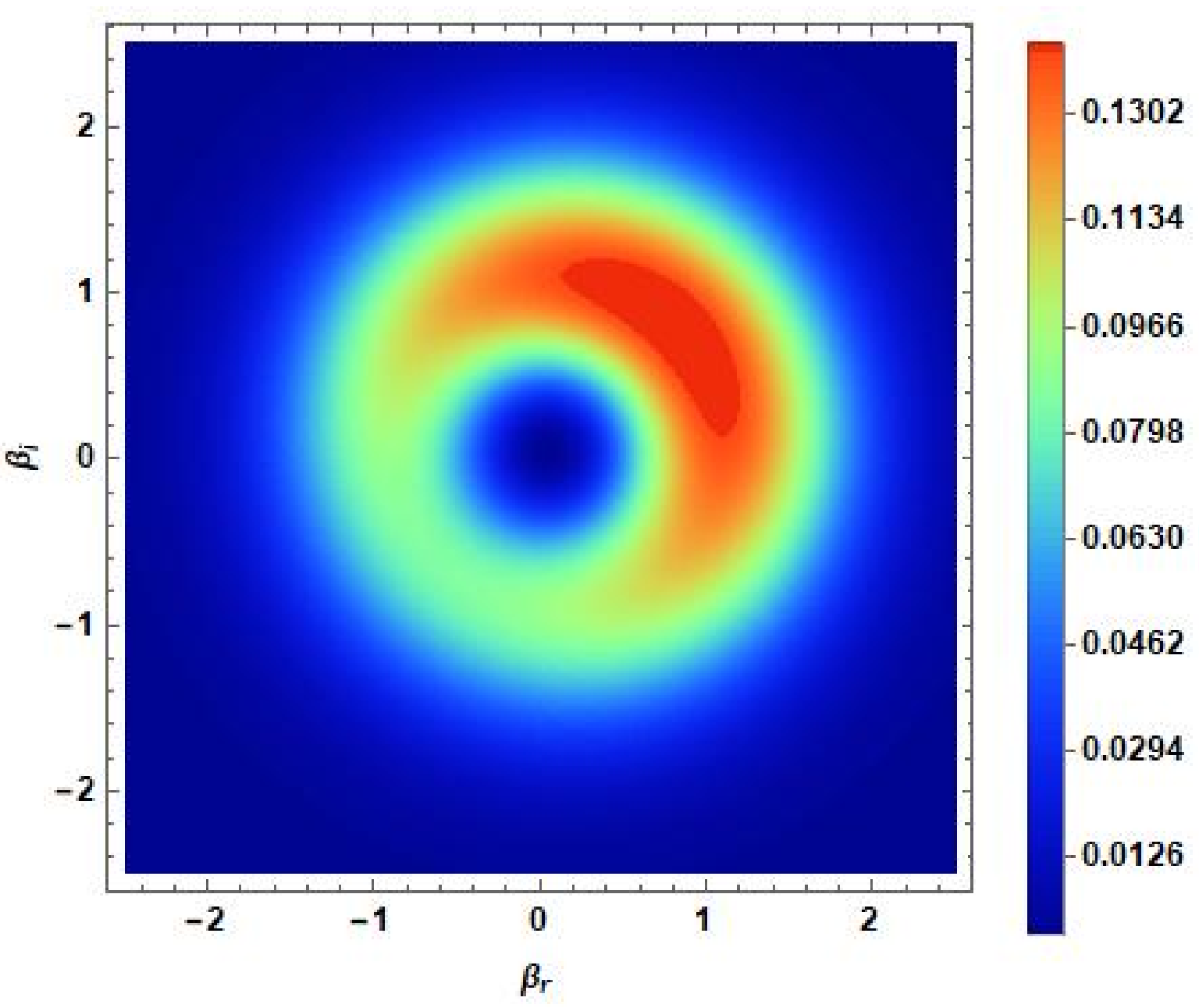}} \subfigure[]{\includegraphics[scale=0.3]{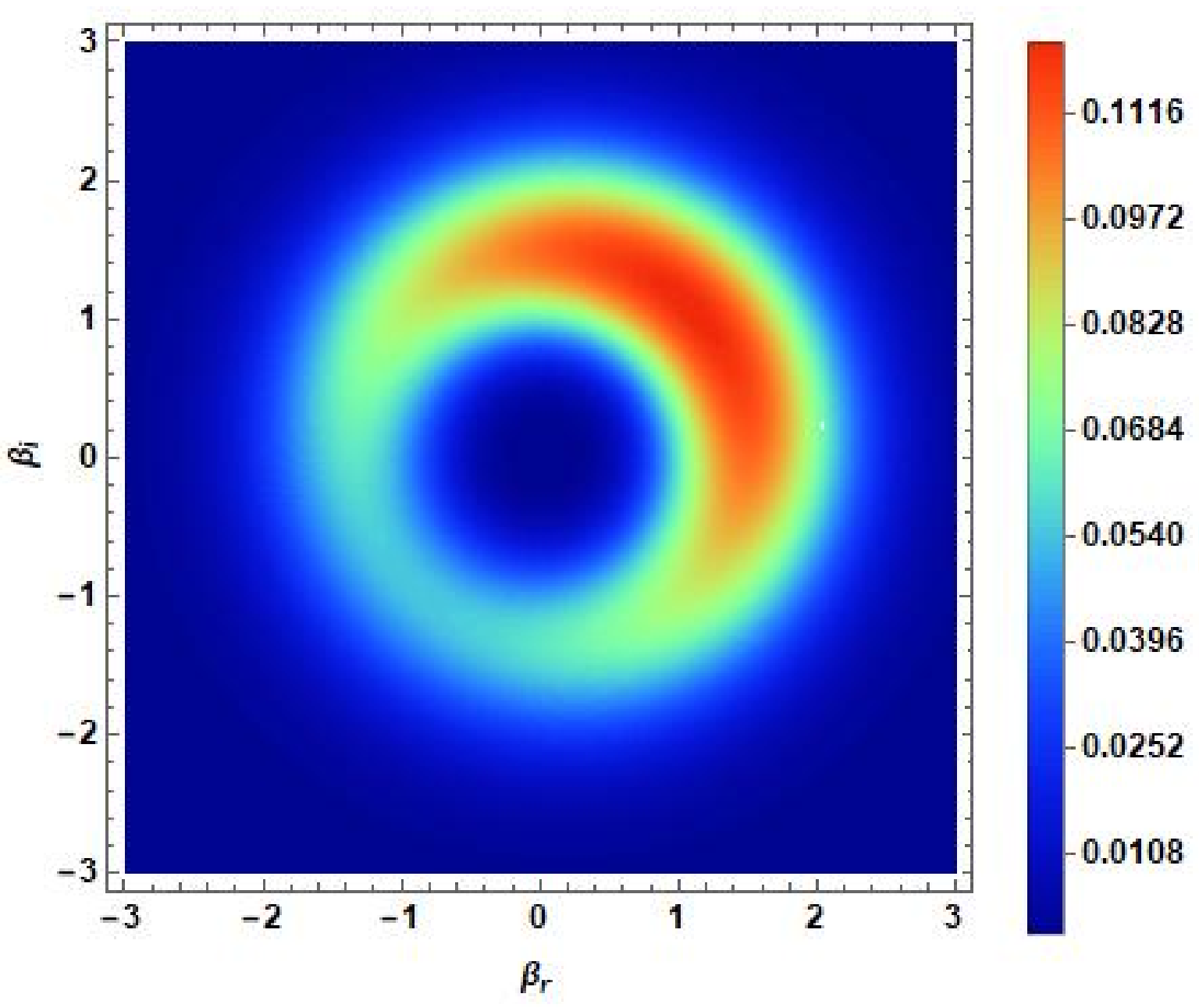}}
\par\end{centering}
\caption{\label{fig:Q function} (Color online) $Q$ function for PASDFS $|\psi\left(k,q,n,\alpha\right)\rangle$
with (a) $k=q=n=1,$ (b) $k=2,\,q=n=1,$ and (c) $q=2,\,k=n=1$ with
$\alpha=\frac{1}{5\sqrt{2}}\exp\left(\iota\pi/4\right)$. (d) Similarly,
$Q$ function of PASDFS with $q=2,\,k=n=1$ and $\alpha=\sqrt{2}\exp\left(\iota\pi/4\right)$.
$Q$ function for $|\psi\left(k,q,n,\alpha\right)\rangle$ with (e)
$k=q=1,\,n=2$ and (f) $q=1,\,k=n=2$ for $\alpha=\frac{1}{5\sqrt{2}}\exp\left(\iota\pi/4\right)$.}
\end{figure}

\section{Quasidistribution function: $Q$ function \label{sec:Qfn}}

Quasidistribution functions allow us to calculate the average values
of an operator analogous to classical phase space and are also witnesses
of nonclassicality. For instance, negative values of Wigner and $P$
functions and zeros of $Q$ function are signatures of nonclassicality
present in the quantum state \cite{agarwal2013quantum,thapliyal2015quasiprobability}.
All these quasidistribution functions can be defined in terms of each
other and are also related to the corresponding characteristic function.
Therefore, one can also use the quasidistribution functions to study
non-Gaussianity both qualitatively and quantitatively \cite{ivan2012measure}.
Non-Gaussian states are also found useful in a set of secure quantum
communication schemes \cite{lee2019quantum,srikara2019continuous,borelli2016quantum}. 

Here, we discuss only $Q$ function of an arbitrary state $\rho$
which can be defined as \cite{husimi1940some}

\begin{equation}
Q=\dfrac{1}{\pi}\langle\beta|\rho|\beta\rangle,\label{eq:Q-function}
\end{equation}
where $|\beta\rangle$ is a coherent state. The analytic form of $Q$
function for PASDFS can be expressed as

\begin{equation}
\begin{array}{ccc}
Q & = & \frac{N^{2}}{\pi}\exp\left[-\left|\beta\right|^{2}\right]\left|\sum\limits _{m=0}^{\infty}C_{m}\left(\alpha,n,k,q\right)\frac{\beta^{*\left(m+k-q\right)}}{\sqrt{\left(m+k-q\right)!}}\right|^{2}.\end{array}\label{eq:Q-pasdfs}
\end{equation}
Here, We will establish non-Gaussianity inducing behavior of photon
addition and Fock parameter (cf. Fig. \ref{fig:Q function}), which
are so far illustrated as nonclassicality inducing and phase altering
operations. Clearly, with photon addition tendency of quasidistribution
away from Gaussian behavior is visible, while with photon subtraction
squeezing along particular phase angle chosen by displacement parameter
can be observed. This squeezing can be noticed to be more appreciable
for higher values of displacement parameter (cf. Fig. \ref{fig:Q function}
(c)-(d)). From Fig. \ref{fig:Q function} (e)-(f), it can be observed
that Fock parameter and photon addition have a similar effect in the
phase space. As zeros of $Q$ function are signature of nonclassicality,
PASDFS shows nonclassicality in Fig. \ref{fig:Q function} (b), (e),
and (f). 

\section{Conclusions \label{sec:Conclusions}}

Finally, we would like to throw some light on the nonclassical behavior
of PASDFS using different witnesses of lower- and higher-order nonclassicality.
The significance of this choice of state is its uniqueness that a
class of engineered quantum states can be achieved as the reduced
case of PASDFS $|\psi\left(k,q,n,\alpha\right)\rangle$, like photon
added DFS $\left(q=0\right)$, photon subtracted DFS $\left(k=0\right)$,
DFS $\left(q=k=0\right)$, PACS $\left(n=q=0\right)$, photon subtracted
coherent state $\left(n=k=0\right)$, coherent state $\left(n=k=q=0\right)$,
and Fock state $\left(n=k=q=\alpha=0\right)$. Some of the reduced
states have been experimentally realized and in some cases optical
schemes for generation have been proposed, so this family of states
is apt for various challenging tasks to establish quantum dominance.
The state under consideration requires various non-Gaussianity inducing
quantum engineering operations and thus our focus here was to analyze
the relevance of each operation independently in the nonclasscial
features (listed in Table \ref{tab:Properties of PASDFS}) observed
in PASDFS. To study the nonclassical properties of PASDFS, a set of
moments-based criteria for Klyshko's, Agrwal-Tara's, and Vogel's criteria,
as well as lower- and higher-order antibunching, HOSPS, and squeezing.
Further, phase properties for the same state are also studied using
phase distribution function and phase fluctuation. Finally, non-Gaussianity
and nonclassicality of PASDFS is also studied using $Q$ function. 

\begin{table}
\begin{centering}
\begin{tabular}{ccc}
\toprule 
S. No. & Nonclassical Properties & Observed in PASDFS\tabularnewline
\midrule
1 & Lower-order and higher-order Antibunching & yes\tabularnewline
2 & Higher-order sub Poissionian photon statistics & yes\tabularnewline
3 & Lower-order and higher-order squeezing  & yes\tabularnewline
4 & Klyshko's criterion & yes\tabularnewline
5 & Agarwal-Tara's criterion & yes\tabularnewline
6 & Vogel's criterion & yes\tabularnewline
7 & Phase distribution function & -\tabularnewline
8 & Phase fluctuation & yes\tabularnewline
9 & $Q$ function & yes\tabularnewline
\bottomrule
\end{tabular}
\par\end{centering}
\caption{\label{tab:Properties of PASDFS} Summary of the nonclassical properties
of PASDFS. }
\end{table}
The present study reveals that with an increase in the order of nonclassicality
the depth of nonclassicality witnesses increase. Additionally, higher-order
nonclassicality criteria were able to detect nonclassicality in the
cases when corresponding lower-order criteria failed to do so. Different
nonclassical features are observed for smaller values of displacement
parameter, which can be sustained for higher values by increasing
the number of subtracted photon. Photon addition generally improves
nonclassicality, and this advantage can be further enhanced for the
higher (smaller) values of displacement parameter using photon subtraction
(Fock parameter). The HOSPS nonclassical feature is only observed
for the odd orders. As far as squeezing is concerned, only photon
subtraction could induce this nonclassicality. Large number of photon
addition can be used to observe squeezing at higher values of displacement
parameter at the cost of that present for smaller $\alpha$. Photon
subtraction alters the phase properties more than photon addition,
while Fock parameter has an opposite effect of the photon addition/subtraction.
The nonclassicality revealed through phase fluctuation parameter shows
similar behavior as Vogel's criterion. Finally, we have shown the
nonclassicality and non-Gaussianity of PASDFS with the help of a quasidistribution
function, namely $Q$ function. 

We hope the present work focused on the characterization of nonclassicality
in the class of states obtained from PASDFS will be useful in the
applications like non-Gaussian quantum information processing. The
present work can be further extended to quantify the amount of nonclassicality
and non-Gaussianity using different measures.

\textbf{Acknowledgment:} AP, SB and VN acknowledge the support from
Interdisciplinary Cyber Physical Systems (ICPS) programme of the Department
of Science and Technology (DST), India, Grant No.: DST/ICPS/QuST/Theme-1/2019/6.
KT acknowledges the financial support from the Operational Programme
Research, Development and Education - European Regional Development
Fund project no. CZ.02.1.01/0.0/0.0/16 019/0000754 of the Ministry
of Education, Youth and Sports of the Czech Republic.

\bibliographystyle{apsrev4-1}
\bibliography{biblio}

\end{document}